\documentclass[preprint]{aastex}

\begin{document}

\title{YZ Phe: a very short period K-type contact binary with variation of the O'Connell effect and orbital period change}

\author{T. Sarotsakulchai\altaffilmark{1,2,3},
S.-B. Qian\altaffilmark{1,2,4,5},
B. Soonthornthum\altaffilmark{3},
X. Zhou\altaffilmark{1,4,5},
J. Zhang\altaffilmark{1,4,5}, \\
L.-J. Li\altaffilmark{1,4,5},
D. E. Reichart\altaffilmark{6},
J. B. Haislip\altaffilmark{6},
V. V. Kouprianov\altaffilmark{6} and
S. Poshyachinda\altaffilmark{3}}

\altaffiltext{1}{Yunnan Observatories, Chinese Academy of Sciences, 650216 Kunming, P.R. China} \email{huangbinghe@ynao.ac.cn}
\altaffiltext{2}{University of Chinese Academy of Sciences, 19 A Yuquan Rd., Shijingshan, 100049 Beijing, P.R. China}
\altaffiltext{3}{National Astronomical Research Institute of Thailand, Donkaew, Maerim, Chiang Mai 50180, Thailand}
\altaffiltext{4}{Key Laboratory of the Structure and Evolution of Celestial Objects, Chinese Academy of Sciences, 650216 Kunming, P.R. China}
\altaffiltext{5}{Center for Astronomical Mega-Science, Chinese Academy of Sciences, 20A Datun Rd., Chaoyang District, 100012 Beijing, P.R. China}
\altaffiltext{6}{Department of Physics and Astronomy, University of North Carolina, CB \#3255, Chapel Hill, NC 27599, USA}


\begin{abstract}
YZ Phe is a very short-period contact binary (Sp.= K2 V) with an orbital period of 0.2347 days near the short period limit (0.22\,d). We present the complete light curves in $VRI$ bands, which photometric data were obtained with the 0.61-m reflector of PROMPT-8 at CTIO in Chile during June to October 2016 and August 2017. The photometric solutions were determined by using the W-D method and the results reveal that YZ Phe is a W-subtype shallow contact binary ($f\sim$ 10\%, $q$ = 2.635 or $1/q$ = 0.379 for W subtype) with rotational motion of a large hot spot on the more massive component, showing a strong O'Connell effect with variation of maxima in photometric time series at period of 4.20 yr and stellar cycle at period of 1.28 yr. By compiling all available eclipse times, the result shows a long-term period decrease at a rate of $\mathrm{d}P/\mathrm{d}t = -2.64(\pm 0.02)\times 10^{-8}$ d $yr^{-1}$, superimposed on a cyclic variation ($A_3$ = 0.0081 days and $P_3$ = 40.76 years). This variation cannot be explained by Applegate mechanism. Thus, the cyclic change may be interpreted as light-travel time effect via the presence of a cool third body. Based on photometric solutions, the third light was detected with 2\% of total light in V and I bands. Those support the existence of a third body. For the long-term period decrease, it can be explained by mass transfer from the more massive component ($M_2 \sim 0.74 M_{\odot}$) to the less massive one ($M_1 \sim 0.28 M_{\odot}$) or plus AML via magnetic braking. With $1/q$ $<$ 0.4 and long-term period decrease, all suggest that YZ Phe is on the AML-controlled state and its fill-out factor will increase, as well as the system will evolve into a deeper normal contact binary.
\end{abstract}

\section{Introduction}
Study the late evolutionary stage of low-mass contact binary systems is important to understand their physical properties and stellar evolution because their formation and evolution are still an open question in astrophysics. K and M type contact binaries are expected to have a very short period (P $<$ 0.3\,d). A few of K-type systems are very close to the short-period limit (P $\sim$ 0.22\,d). This kind of binaries are very rare, which makes them are an important system to investigate the cause of period cut-off. The explanations for contact binaries which period near the period cutoff have been widely discussed by many authors (e.g. Rucinski 1992, 2007; Stepien 2006a, 2011; Nortion et al. 2011; Jiang et al. 2012). To date, there are numerous eclipsing binaries below the period limit which have been discovered and many of them were investigate well i.e., M-type systems and most of them are found to have a minimum period below 0.22\,d. For example, M dwarf + M dwarf; white dwarf (WD) + M dwarf binaries (Drake 2014a, Qian 2015a). However, the observation and investigation of K-type contact binaries are still very rare, especially the systems with period shorter than 0.3\,d as explained by Bradstreet (1985). A few K-type contact binaries that have been studied e.g. CC Com (P $\sim$ 0.2207\,d, Yang et al. 2009a); BM UMa (P $\sim$ 0.2712\,d, Yang et al. 2009b); FY Boo (P $\sim$ 0.2411\,d, Samec et al. 2011); BI Vul (P $\sim$ 0.2518\,d, Qian et al. 2013); V1799 Ori (P $\sim$ 0.2903\,d, Liu et al. 2014); 1SWASP J064501.21 (P $\sim$ 0.2486\,d, Liu et al. 2014); V1104 Her (P $\sim$ 0.2279\,d, Liu et al. 2015); NSVS 2701634 (P $\sim$ 0.2447\,d, Martignoni et al. 2016). Some of short period contact binaries which very near the limit (P $\sim$ 0.20 - 0.22\,d) are showing the period decreases and some are showing the period increases as found in period change measurements of 143 SuperWASP eclipsing binaries by Lohr et al. (2013). Moreover, the investigation of SuperWASP J015100.23 (Qian et al. 2015b) as a contact binary below the period limit (P=0.2145\,d) suggests that contact binaries below the limit are not rapidly destroyed. But in another case e.g. NSVS 925605 (Dimitrov \& Kjurkchieva 2015) which exhibits high activity including strong H$\alpha$ emission line, large cool spots and high fill-out factor (0.7) that may be a progenitor of the predicted merging. Many candidates of mergers exhibit a strong magnetic activity, but the relationship between high fill-out factor for a merger and its magnetic activity is unknown. The relation between the short period contact binaries around the period cutoff and their period changes which investigated from the SuperWASP data are still unclear. This may be because of the lack of times of light minimum which were obtained only from the SuperWASP archive. Long-term monitoring and new data from all available surveys are needed to investigate period change. 

In addition, those short-period contact binaries also provide a good chance to investigate their magnetic activities and the O'Connell effect (O'Connell 1951) that expected to occur on late-type stars e.g. F, G, K and M types. In single stars, there are many studies about magnetic cycle e.g., Scargle (1982); Wright et al. (2011); Suarez Mascareno et al. (2016); Wargelin et al. (2017). They pointed out that there are two periodic signals e.g., the rotational period and stellar cycle. Many late-type single stars were found to have brightness periodic changes in photometric time series and these photometric modulations were found to be induced by magnetic cycles or stellar rotation with spots on host stars. The rotation period and magnetic cycle can distinguish clearly, for example, the statistics of stellar cycle length and rotation period by Suarez Mascareno et al. (2016), i.e., F-type stars (stellar cycle is 9.5 yr and rotation period is 8.6 d) and K-type stars (stellar cycle is 8.5 yr and rotation period is 27.4 d). The other single star e.g. G-type HD 2071 (stellar cycle is 11.3 yr with semi-amplitude of 4.9 mmag and rotation period is 29.6 d with semi-amplitude of 3.1 mmag), K-type HD 224789 (stellar cycle is 7.4 yr with semi-amplitude of 6.7 mmag and rotation period is 16.6 d with semi-amplitude of 6.8 mmag), Proxima Centauri (stellar cycle is 6.8 yr and rotation period is 83.2 d). For light curve variations or the O'Connell effect in close binaries, many models have been proposed to explain: (1) the presence of spots ; (2) the mass transfer between the component with long-term orbital period change; (3) asymmetric circumfluence from Coriolis forces, see Koju et al. (2015). However, the magnetic activities in eclipsing binary stars are more complicated than the case of single stars because of the movement of spots on one or both components along with the rotation of each star. For example, the study on Kepler contact binaries by Tran et al. 2013 had shown that contact binaries with orbital period of 0.2 to 0.5 days have a strong behavior of anti-correlations between the primary and secondary minima (min-I and min-II) in the $O-C$ curves and this behavior was also found in the $O-C$ curves of the primary and secondary maxima (max-I and max-II). They found that the orbital period change via mass transfer cannot explain the anti-correlated behavior, the spot movement in longitude direction is the plausible reason to cause the phase shifts or anti-correlation between the minima or maxima in the $O-C$ curves. The $O-C$ diagrams of active contact binaries with short timescale changes, high frequency and low amplitude perturbations were found to be induced by starspots, e.g. Kalimeris et al. (2002), who explained that cool spots can affect the eclipse times. This may result in a difference between real orbital period changes and the observed $O-C$ values in active contact binaries which modulated by spots. Other study on longitudinal motion of spots in Kepler binaries had been more investigated by Balaji et al. 2015, who employed the same method by Tran et al. (2013) to track the rotational movements of spots. They also found that the motions of spots are related to differential rotation of the host stars. As the result, there are many factors that may cause the variations in the $O-C$ curves of light minima in close binaries. Not only the effects from spots, but also the magnetic activity i.e. Applegate mechanism (Applegate 1992) which supported by Lanza \& Rodono (1999), they proposed that the quasi-periodic perturbations in the $O-C$ curves are the effect from quadrupole moment variations of magnetic cycle. In another aspect, Borkovits et al (2015, 2016) had also proposed that the eclipse time variations (ETVs) in the $O-C$ curves are resulted from the dynamical effects of additional companions i.e., a third body in a binary system or the light-travel time effect (LTTE) through the existence of a tertiary component (Irwin 1952). This suggestion is in agreement with theoretical study by Eggleton \& Kiseleva-Eggleton (2001, 2006) that most of short period close binaries are three-body systems, where the third body can help the central binary to become shorter orbital period or shrink via Kozai cycles (Kozai 1962) and tidal friction. However, the properties of the magnetic activity in close binaries are still unclear (Qian et al. 2014) and needed to be investigated.

Recently, a study of period distribution from LAMOST spectral survey of EW-type binaries by Qian et al. (2017) has shown that the typical period of EW-type binary stars is about 0.29\,d. Based on the LAMOST survey, it is also found that there is a sharp cut-off at 0.2\,d in EW-type binaries which is similar to several surveys by other investigators (e.g. Paczynski et al. 2006; Drake et al. 2014b, 2017), while a publication from the Wide-field Infrared Survey Explorer (WISE) by Chen et al. (2018) reveals a minimum period of 0.19\, d for EW-type binaries. Norton et al. (2011) also found 53 W UMa type binary candidates in SuperWASP data with periods close to the short-period limit. Besides, a study of the light-travel time effect (LTTE) in short period eclipsing binaries (EBs) by Li et al. (2018) suggests that contact binaries with periods close to 0.22\,d are commonly accompanied by relative close tertiary companions. The frequency of EBs with tertiary companions in their sample increases as the period decreases, especially contact binaries with period $<$ 0.26\,d. This indicates that low mass and short period contact binaries should be found to have a cyclic change in their $O-C$ analysis with LTTE via the presence of a third body orbiting around their host binary stars. This is in agreement with the suggestion by Qian et al. (2013, 2015a) that a third body may have driven a binary to shorter orbital periods and evolve to contact phase. Namely, the formation of low-mass contact binary is driven by AML via magnetic braking, but the timescale of the AML for this system is long. However, there are many contact binaries that have been found to have companions (e.g., D'Angelo et al. 2006; Pribulla \& Rucinski 2006; Duerberk \& Rucinski 2007; Rucinski et al. 2013), which are a plausible reason to explain that some W UMa type binaries were formed by Kozai cycle (e.g., Eggleton \& Kiseleva 2001; Paczynski et al. 2006). Thus, the close-in companion should play an important role for the formation and evolution by removing angular momentum from the central binary via Kozai cycles (Kozai 1962) during the early dynamical interaction or late evolution which cause the binary to have a very low angular momentum and a very short initial orbital period. In this way, the initially detached binaries with period less than 1\,d will fill their primaries Roche lobes with stable mass transfer and evolve toward contact configuration via AML from magnetic stellar wind as discussed by many authors (e.g., Bradstreet \& Guinan 1994; Yakut \& Eggleton 2005; Stepien 2006b; Qian et al. 2017; Qian et al. 2018). However, if the short period contact binaries near the period limit are likely to show the period decrease, the separation between the components is closer and they will be unstable to maintain the contact configuration and with high degree of contact, ultimately the two components will merge into single rapidly rotating star like the case of V1309 Sco (e.g. Tylenda 2011) as explained by Lohr et al. (2012). All these reasons make the short-period contact binaries, with period close to 0.22\,d, important targets to understand formation and evolution of such binaries, as well as their interaction with additional companions.

YZ Phe (S7172) is a high galactic southern latitude contact binary, which known as a very short period binary with 0.2347 days, and it was discovered by Hoffmeister (1963). Later, photometric data were obtained by Gessner \& Meinunger (1974 and 1976) with seven new times of minimum light and its determined period of 0.3052 days. In 1989, $UBVRI$ photometric observations were published by Jones (1989), who determined standard magnitudes of YZ Phe for Min I and Max II. He found that the $V-R$ and $V-I$ colors implied a slightly later type (K2-K5) than $B-V$ and $U-B$ type (K0-K2). Moreover, he concluded that YZ Phe was a W-subtype contact binary and stated that the value of 0.3052 days was incorrect because the period given by Gessner \& Meinunger was a 1 day alias of the true period at that epoch, 0.2338 days. At the same year, the more observations were performed by Kilkenny \& Marang (1990), who transferred all data to the $UBV(RI)_{C}$ standard system and they also got nine new times of minimum light. They found that the phased light curve has O'Connell effect with clearly unequal maxima and minima. By applying a phase-dispersion minimisation technique, a period of 0.234726 $\pm0.000002$ days was derived. The other observations of YZ Phe were done at the same year by Samec et al. (1993) and the complete $BVR_{C}I_{C}$ light curves were obtained. Their period study with four new times of minimum light revealed that its period had remained constant at 0.23472 days for over 30 years. With their early analysis, it was reported that YZ Phe was a W-subtype contact binary with a mass ratio of 0.41, a fill-out of 16\% and $\Delta$T=380 K. Furthermore, their light-curve modeling showed that there was a cool spot on the cooler component with a large $46^{\circ}$ radius and a temperature factor of 0.96. Later, Samec \& Terrell (1995) published new analysis results from first publication in 1993, they found that the more massive component was the cooler one. From their new light-curve modeling, it revealed that it strongly favors the presence of a large $54^{\circ}$ radius superluminous region instead of a dark spot region, which usually invoked to model asymmetries on convective contact binaries. In addition, the lowest residual of the hot spot model solution with a temperature factor of 1.03, indicates that the system consists of a K3\,V primary and a K2\,V secondary component with a mass ratio of $\sim$0.40 and shallow contact with a low fill-out factor of 2\%, while the temperature $\Delta$T=255 K. Besides, new full light curve of YZ Phe in V-band was published by Irawati et al. (2013). As their preliminary results indicated that both primary and secondary minima were shifted by 0.1 phase. The shift indicated that the ephemeris they used from Kreiner (2004) is inaccurate to predict the eclipse time and the orbital period of YZ Phe has changed since the last ephemeris measurement.

In this paper, we report new results from photometric observations in V, R and I bands for YZ Phe. The orbital period changes will be re-investigated and the existence of a third body will be discussed. Additionally, the long-term changes in Maximum I (the primary maximum) and Maximum II (the secondary maximum)  from photometric time series, as well as the variations of the difference between the two maxima (Max I minus Max II) from high cadence photometry of SuperWASP data will be examined.

\begin{table}
\scriptsize
\caption{Coordinates of YZ Phe, the comparison, and the check stars.}
\begin{center}
\begin{tabular}{lcccccccc}\hline
Targets &$\alpha_{2000}$ &$\delta_{2000}$ &$B$&$V$&$R$&$J$&$H$&$K$\\
\hline
YZ Phe &$01^{\textrm{h}}42^{\textrm{m}}25^{\textrm{s}}.9$ & $-45^\circ57^{\prime}03^{\prime\prime}.9$ &13.91&12.96&12.10&11.321&10.765&10.708\\
UCAC2 12136273 (Com)& $01^{\textrm{h}}42^{\textrm{m}}24^{\textrm{s}}.3$ & $-45^\circ54^{\prime}06^{\prime\prime}.5$ &12.35&11.72&11.61&10.465&10.132&10.096\\
UCAC3 89-3647 (Chk)& $01^{\textrm{h}}42^{\textrm{m}}29^{\textrm{s}}.7$ & $-45^\circ52^{\prime}47^{\prime\prime}.6$ &13.74&12.78&12.48&10.921&10.475&10.310\\
\hline
\end{tabular}
\end{center}
\end{table}

\section{Photometric Observations}

\begin{figure}
\begin{center}
\includegraphics[angle=0,scale=0.5]{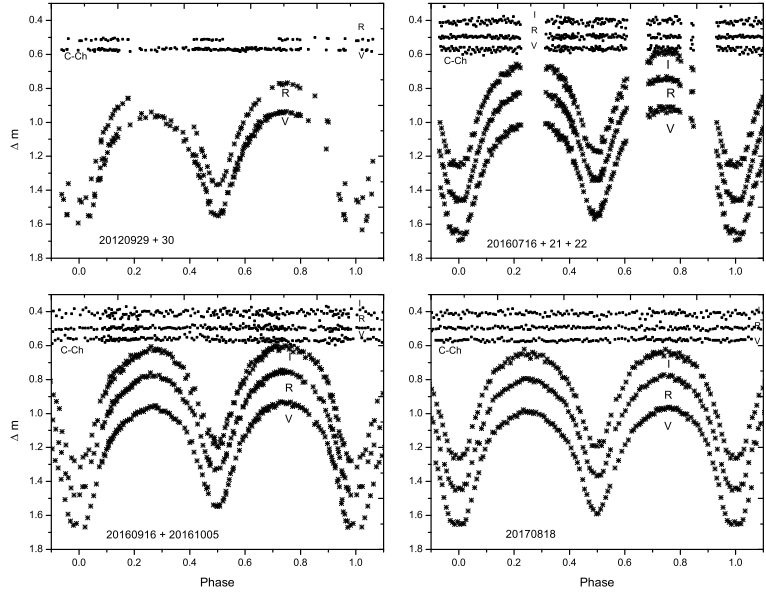}
\caption{The first set of multi-color CCD light curves in V, R bands were obtained with PROMPT-5 in September 2012 (top-left panel) and the other sets in VRI bands were obtained during July, September-October 2016 and August 2017 with PROMPT-8. The differential magnitude between the comparison and the check stars (C-Ch) are stable without significant variation as shown in the figure.}
\end{center}
\end{figure}

The main photometric observations of YZ Phe in $V$, $R$ and $I$ bands were obtained from June to October 2016 and August 2017, with back illuminated Apogee F42 2048$\times$2048 CCD photometric system, attached to the 0.61-m Cassegrain reflecting telescope of PROMPT-8\footnote{PROMPT-8 is the Thai Southern Hemisphere Telescope (TST), operated in collaboration between National Astronomical Research Institute of Thailand (NARIT) and the University of North Carolina (UNC) at Chapel Hill in a part of the UNC-led PROMPT project, http://skynet.unc.edu.} robotic telescope. The other observations were performed with PROMPT-5, by the courtesy of data from Dr. Puji Irawati (Irawati et al. 2013), in 29 and 30 September 2012. Both telescopes are located at the Cerro Tololo Inter-American Observatory (CTIO) in Chile. The web-based SKYNET client allowed us to request and retrieve image remotely via the internet. It also provided nightly calibration images, including bias, dark, and flat-field images (Layden et al. 2010). The CCD reduction and aperture photometry were done with standard procedure packages of IRAF\footnote{The Image Reduction and Analysis Facility (IRAF), http://iraf.noao.edu.}. The details of variable stars, the comparison and check stars are listed in Table 1. The complete photometric light curves are displayed in Figure 1; the top left panel for September 2012, top right panel for July 2016, bottom left panel for September - October 2016 and bottom right panel for 18 August 2017, respectively. Additionally, we also put all light curves from PROMPT-8 together and use the differential magnitudes of the C-Ch star to check the variations in light curves as shown in Figure 2. The figure shows that the Maximum I and Maximum II of light curves vary from time to time, and the Maxima I are often lower than the Maxima II as the same pattern in paper by Samec \& Terrell (1995).

\begin{figure}
\begin{center}
\includegraphics[angle=0,scale=0.4]{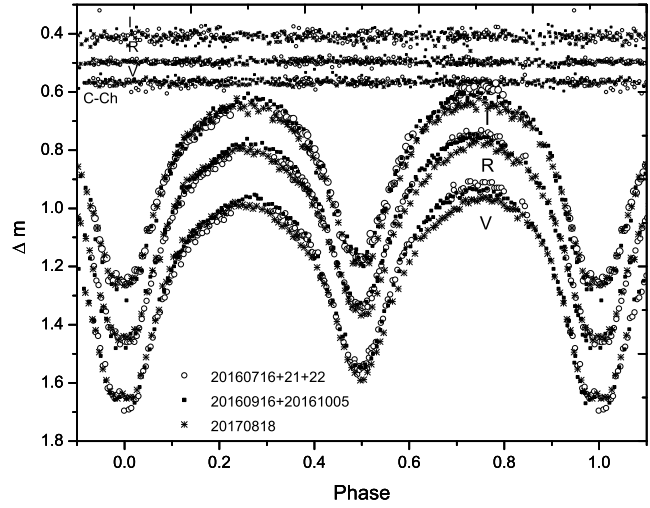}
\caption{All sets of multi-color light curves from PROMPT-8 during 2016 and 2017 in Fig. 1 are put together. The differential magnitudes of the comparison and the check stars (C-Ch) are used to check the variations. The light curves show the significant variations between Max I and Max II, also Min I and Min II. It is also shown that all light curves do not overlap well, indicating that the light curves are variable.}
\end{center}
\end{figure}

\section{The O'Connell effect and magnetic cycle}

According to the photometric solutions by Samec \& Terrell (1995), YZ Phe showed a large O'Connell effect in the light curves which they explained as a large hot spot with radius of 54 deg on the cooler and more massive component. How a large hot spot occurs on the contact binary star is still unknown. In addition, Figure 2 shows the light curve variations and the largest difference between Max-I and Max-II from the light curves in July 2016 is 0.06 mag where Max I is the lowest and Max II is the highest. While, Max I and Max II in light curves of August 2017 is not much different and Max I is still lower than Max II (known as negative O'Connell effect). We compile all photometric data of YZ Phe from other archives including the All-Sky Automated Survey (ASAS; Pojmanski 1997, 2002), the Siding Spring Surveys (SSS; Drake et al. 2017), the SuperWide Angle Search for Planets (SuparWASP; Pollacco et al. 2006), as well as our data from PROMPT-5 and PROMPT-8 to analyze their long-term variation of maxima (Max I and Max II) as shown in Figure 3.  All archives were put together in the same scale to compare each other as displayed in Figure 4. The figure shows clearly a cyclic change with sinusoidal pattern. Therefore, we use a sinusoid function to fit the data and it is found that the period of {\bf cyclic change is 1532.78 $\pm 50.25$ days (4.20 $\pm 0.14$ yr)} with a semi-amplitude of 0.06 $\pm 0.005$ mag. This indicates that the light curves show a long-term variation of maxima that Maxima I or Maxima II vary with time periodically with a maximum magnitude of 0.06 mag or in the other words, the maximum (Max I or Max II) shows a highest magnitude and then it changes to a lowest magnitude, and returns to the highest magnitude again as a cycle. This oscillation of maximum is expected to be due to the rotational motions of spots on one or both components.

\begin{figure}
\begin{center}
\includegraphics[angle=0,scale=0.4]{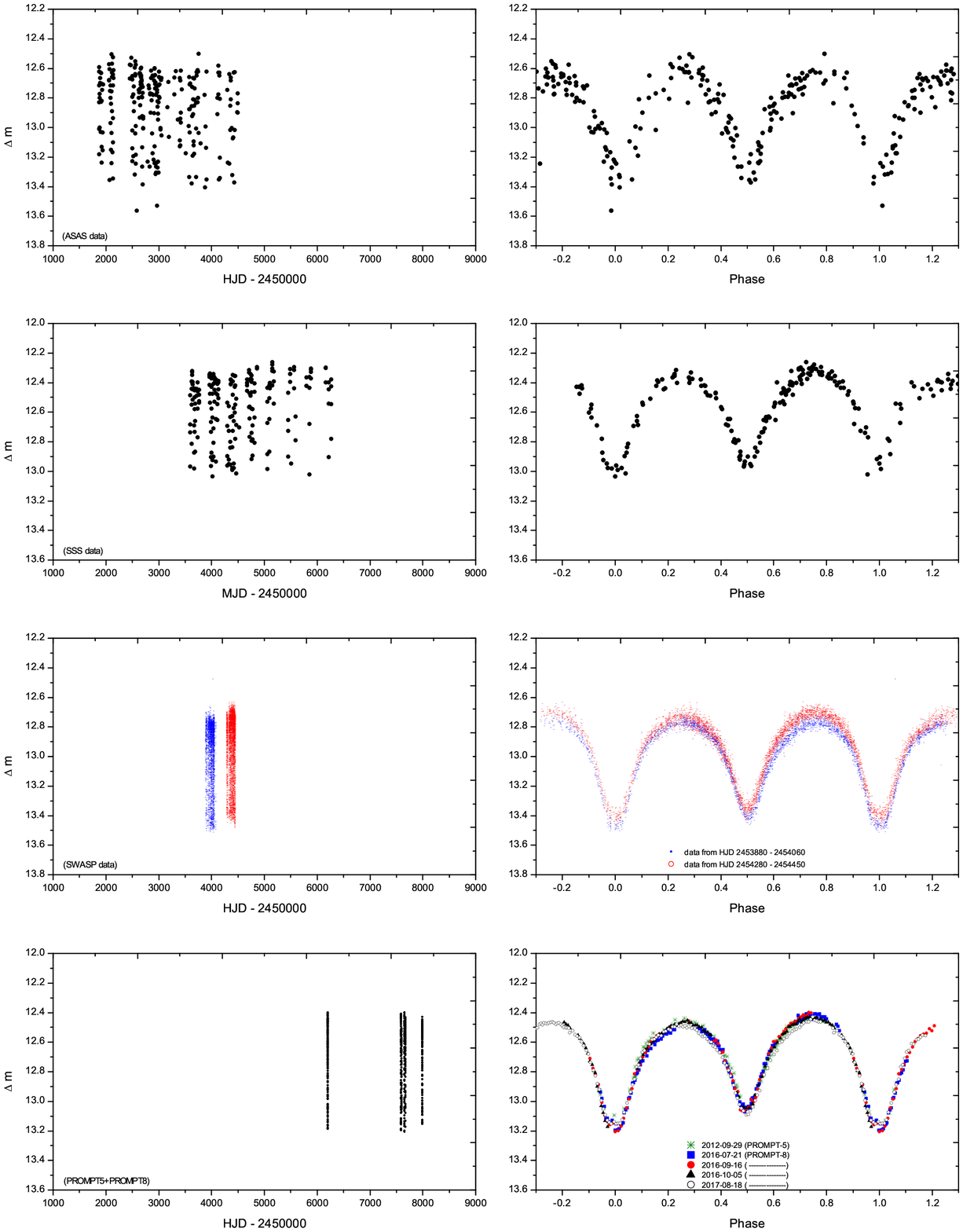}
\caption{All sets of V-band photometric data from ASAS (top panel), SSS (second panel), SuperWASP (third panel) archives, as well as our observations from PROMPT-5 and PROMPT-8 (bottom panel) are put together as function of time. The maxima of all data in left panel show the significant cyclic variations. The right panels show the phased light curves with period of 0.2347 d.}
\end{center}
\end{figure}

\begin{figure}
\begin{center}
\includegraphics[angle=0,scale=0.4]{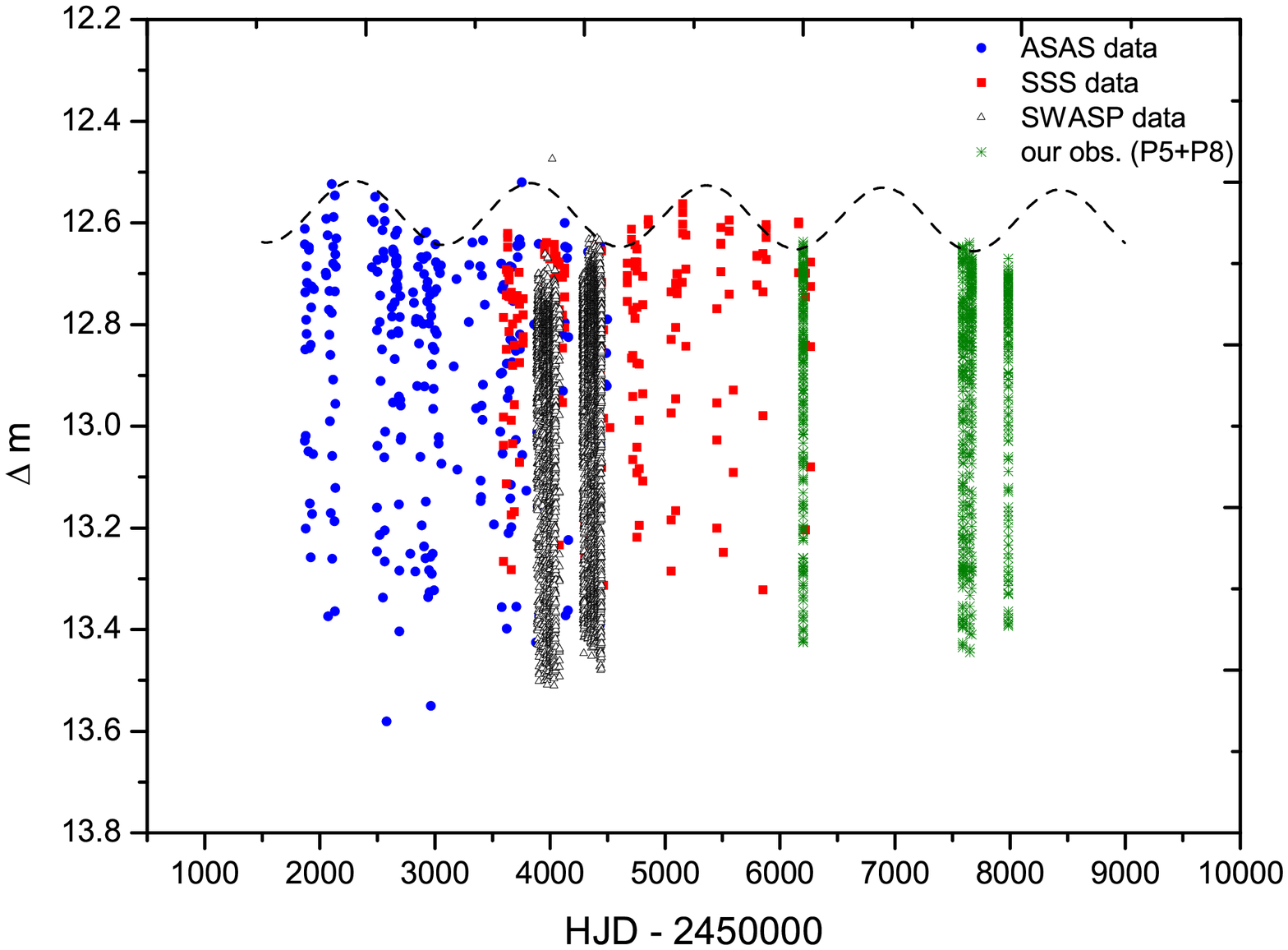}
\caption{All sets of V-band photometric data from Figure 3 are put together in the same scale to check the variation of maxima. The maxima show a long-term variation with sinusoidal best-fit (see dashed line) at {\bf a period of 1532.78 $\pm 50.25$ days (or 4.20 $\pm 0.14$ yr)} with a semi-amplitude of 0.06 $\pm 0.005$ mag.}
\end{center}
\end{figure}

With the SuperWASP data of YZ Phe, we look for details in each light curve and some light curves are shown in Figure 5. We found the variations of maximum. Thus, we use least-squares method to determine each Max I and Max II in each light curve. All differences between Max I and Max II (Maximum I minus Maximum II) are plotted with time (HJD), the results are shown in Figure 6. The observed data are put in the figure and a sinusoidal curve can fit well with both data sets. This reveals that the differences of maximum I and maximum II vary with time as function of sinusoid with a {\bf period of 466.16 $\pm 25.62$ days (1.28 $\pm 0.07$ yr)} and a semi-amplitude of 0.048 $\pm 0.005$ mag. It may suggest that the O'Connell effect of YZ Phe is variable as spot activities or stellar cycle with a period of 1.28 yr and average peak of 0.048 mag. Its stellar cycle is shorter than the period of spot cycle of the Sun (11 yr solar cycle). This means that the components of the contact binary rotate rapidly and rotate faster than the Sun. Sometime its Max I is lower than Max II (negative O'Connell effect) and sometime its Max I is higher than Max II (positive O'Connell effect) with a period of 1.28 yr. These also suggest that the O'Connell effect is variable with two stages of activity; active stage (unstable light curve) where Max I $\neq$ Max II and inactive stage (stable light curve) where Max I = Max II as discussed by Qian et al. (2014). This may involve with the evolution of spots. However, no optical flares were found during the active stage for YZ Phe.

\begin{figure}
\begin{center}
\includegraphics[angle=0,scale=0.4]{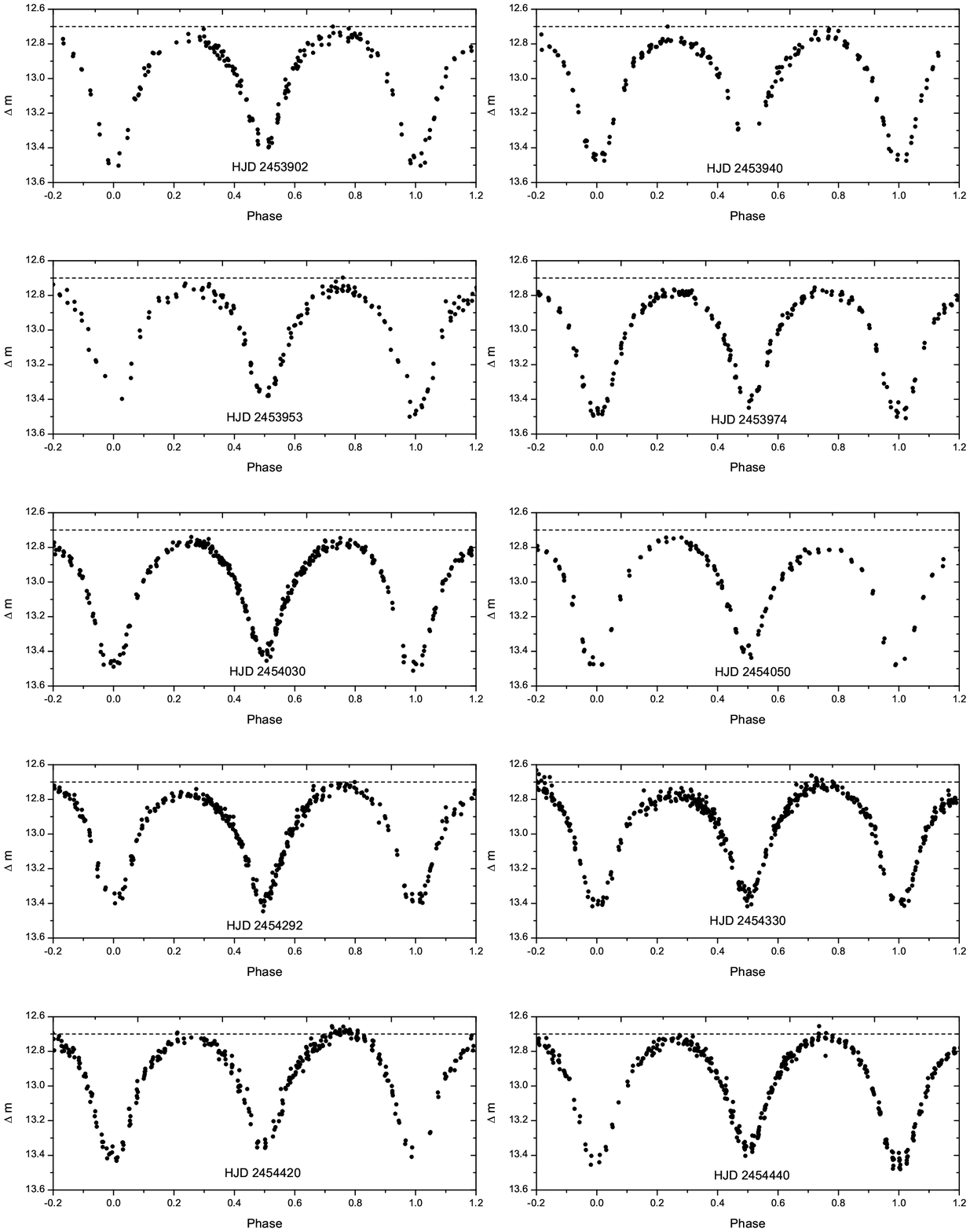}
\caption{Some V-band light curves from SuperWASP archival data (2006 - 2007).}
\end{center}
\end{figure}

\begin{figure}
\begin{center}
\includegraphics[angle=0,scale=0.4]{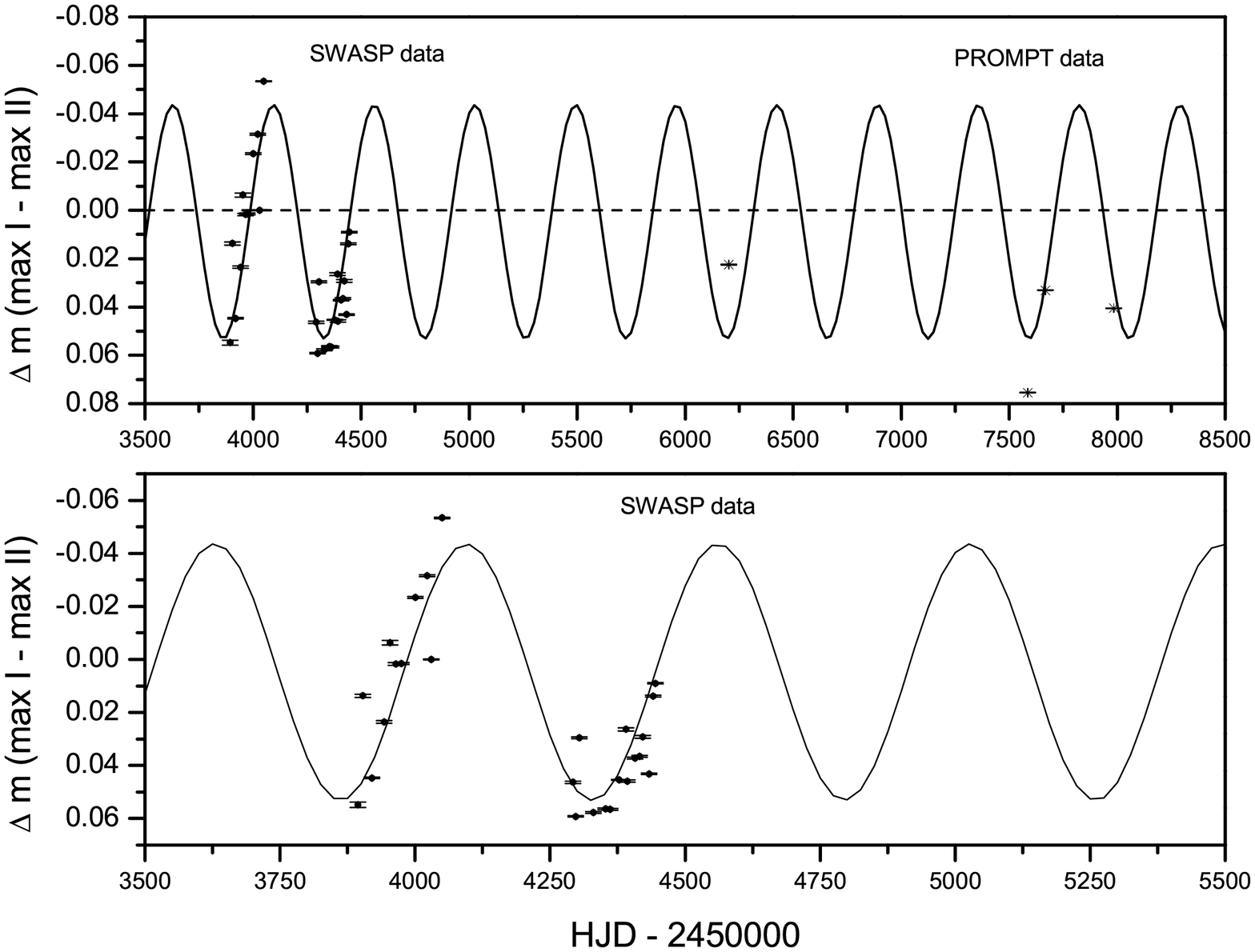}
\caption{Top panel shows a curve fitting for the O'Connell effect from SuperWASP and PROMPT data. It is shown that the variations of difference (Max I minus Max II) can be fitted by sinusoidal curve with a period of {\bf 466.16 days $\pm 25.62$ (or 1.28 $\pm 0.07$ yr)} and semi-amplitude of 0.048 $\pm 0.005$ mag. The dashed line with 0.00 mag means that Max I and Max II are equal without the O'Connnell effect. Bottom panel focuses only on SuperWASP data which were well constrained to the curve fit.}
\end{center}
\end{figure}

\section{Light curve solutions}

The recent publication by Samec \& Terrell (1995) reported that YZ Phe had a large O'Connell effect appearing on light curves which is also seen in our observed light curves as displayed in Figure 1. The light curves obtained in 2012 and July 2016 are not completed and may not be trusted for light curves modeling. Thus, we decided to use the dataset that was obtained in September - October 2016 and the other set in August 2017. Those data were analysed with and without spotted model by using Wilson and Devinney code (Wilson \& Devinney 1971; Wilson 1990, 1994, 2012; Van Hamme \& Wilson 2007) to determine their photometric elements and spot parameters. The third-light contributions are also considered in light-curve synthesis processes.

\begin{figure}
\begin{center}
\includegraphics[angle=0,scale=0.3]{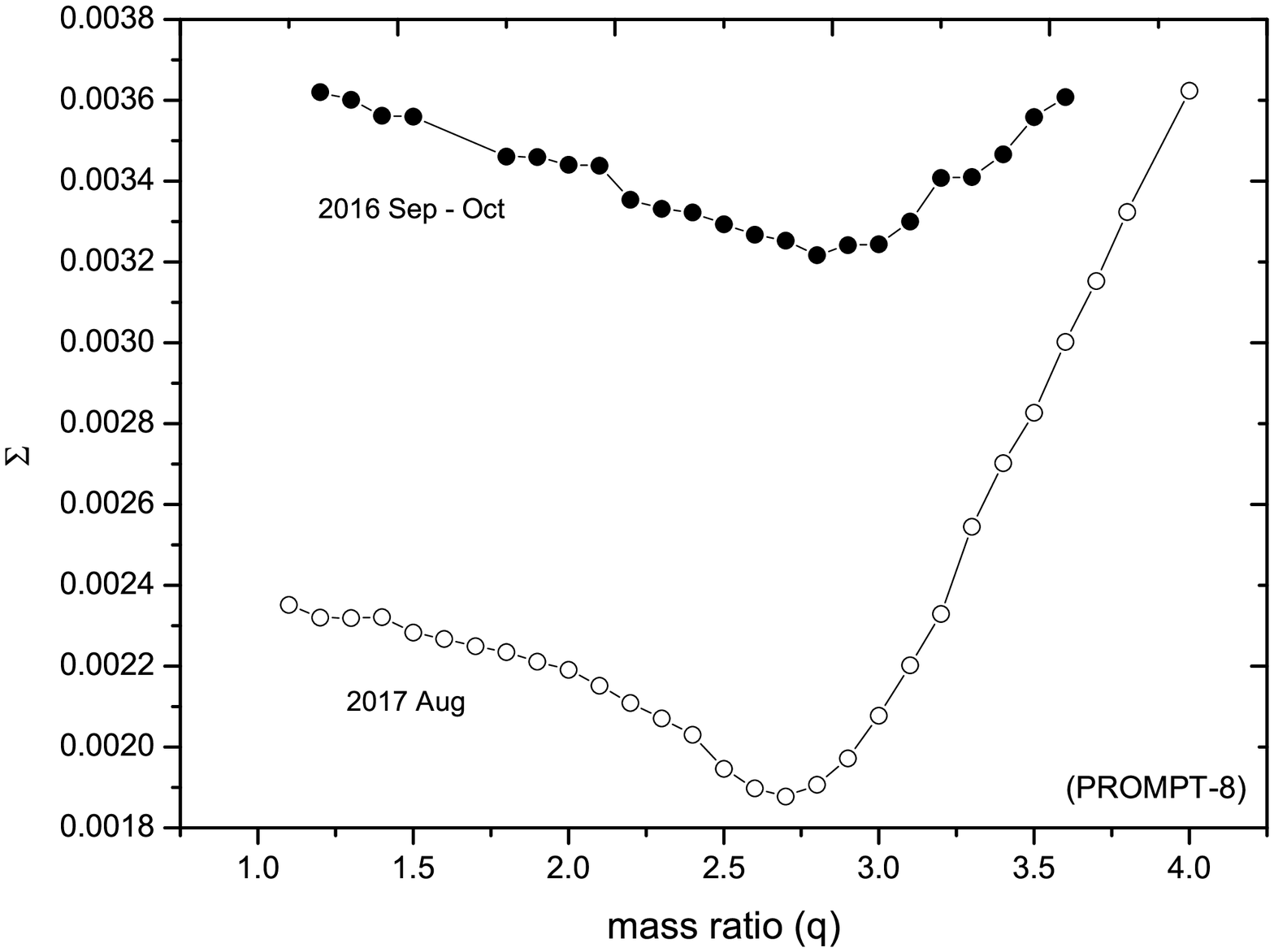}
\caption{Relation Between mass ratio (q) and the sum of weight-squares deviation ($\Sigma)$ for two data sets. The solutions indicate that the optimum mass ratios are in the range between 2.5 to 3.0}
\end{center}
\end{figure}

The spectral classification from Samec \& Terrell (1995) is K3\,V corresponding to $T_{eff}$ = 4800 K (Cox 2000), but AAVSO database reports that $B-V$ = 0.947 which corresponds to a spectral type of K2\,V (Cox 2000) and this agrees with the recent sky survey by {\it Gaia} Data Release 2 (DR2; Gaia collaboration et al. 2016, 2018) that $T_{eff}$ = 4908 K for YZ Phe. Thus, the effective temperature of the primary star ($T_1$) was fixed as 4908 K and assumed that the convective outer layer is already developed for both components. The bolometric albedos for star 1 and 2 were taken as $A_{1} = A_{2}$ = 0.5 (Rucinski 1969) and the values of the gravity-darkening coefficients $g_{1} = g_{2}$ = 0.32 (Lucy 1967) were used. The monochromatic and bolometric limb-darkening coefficients were chosen from Van Hamme's table (Van Hamme 1993) by using logarithmic functions. The adjustable parameters are the inclination ($i$), the mass ratio ($q$), the temperature of Star 2 ($T_{2}$), the monochromatic luminosity of Star 1 ($L_{1V}$, $L_{1R}$ and $L_{1I}$), the dimensionless potential of stars 1 ($\Omega_{1}=\Omega_{2}$ in mode 3 for contact configuration) , respectively.

\begin{figure}
\begin{center}
\includegraphics[angle=0,scale=0.4]{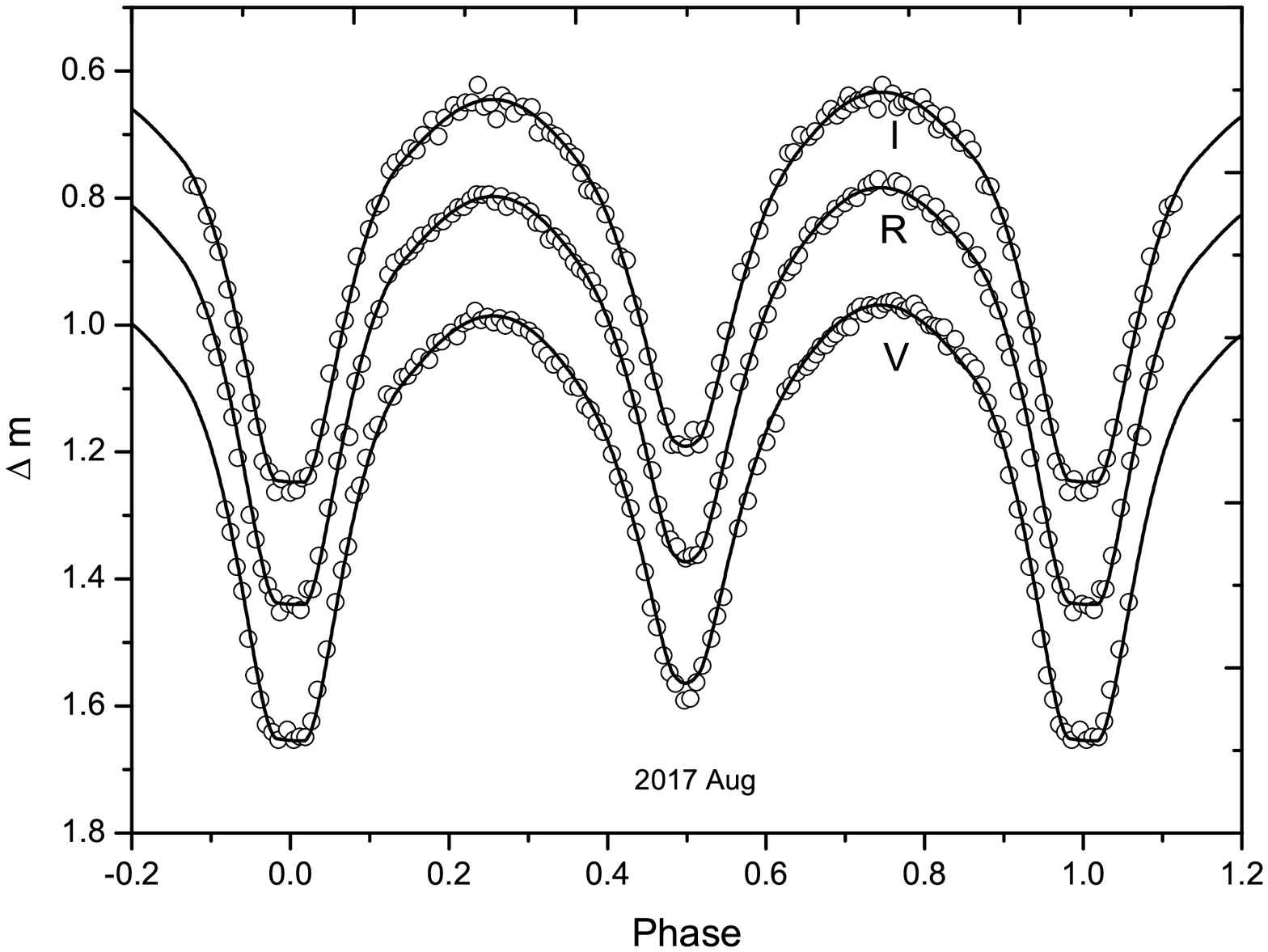}
\caption{Theoretical light curves (solid lines) were computed by using W-D method compared to the observed light curves in August 2017. The results show a small difference between Max I and Max II caused by a large hot spot on the more massive component.}
\end{center}
\end{figure}

In light-curve modeling, the reliable mass ratios are an important parameter and they should be obtained from precise spectroscopic radial velocity measurements. However, no spectroscopic measurements of YZ Phe were performed. Therefore, a $q$-search method is used to determine its initial photometric mass ratio $q_{ph}$ and the mass ratio is set as an adjustable parameter to get a better fit later. Figure 1 indicates that YZ Phe is a W-subtype contact system which its mass ratio $q$ is more than 1.0 ($0 < 1/q < 1.0$) where the more massive star is cooler. During the $q$-search we obtained the initial mass ratio $q$ = 2.8 ($1/q$ = 0.357) for the first set (Sep - Oct 2016) and $q$ = 2.7 (1/q = 0.370) for the second set (August 2017), respectively. The results from $q$-search method are plotted in Figure 7, the solid circles refer to the first set (Sep - Oct 2016) and open circles refer to the second set (August 2017), respectively.

\begin{figure}
\begin{center}
\includegraphics[angle=0,scale=0.4]{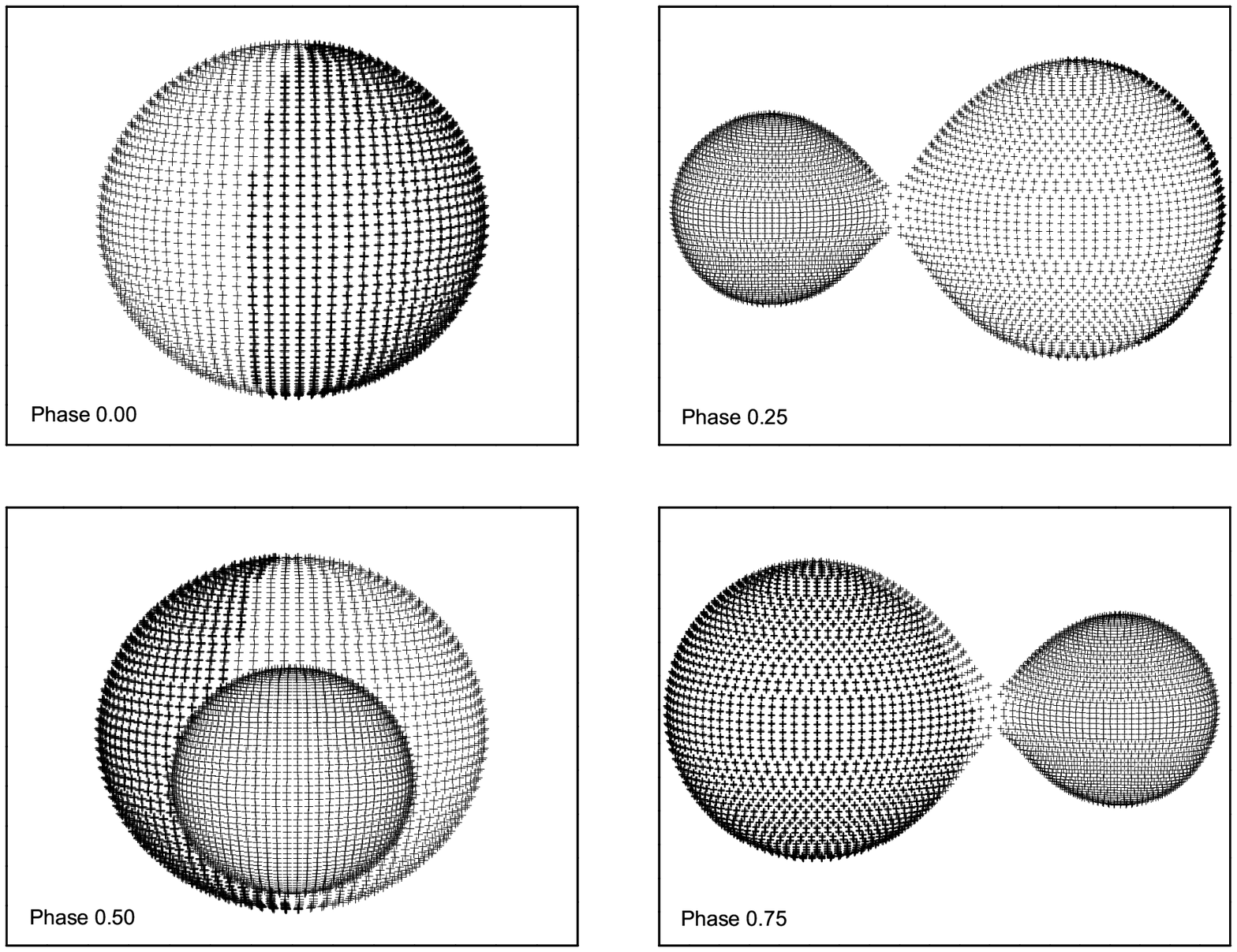}
\caption{The geometrical configurations corresponding to the best-fit of photometric solutions in Figure 8 at phase of 0.00, 0.25, 0.50 and 0.75. The results show a large hot spot (spot radius 89 deg) on the more massive component.}
\end{center}
\end{figure}

Since the light curves in Figure 1 show asymmetry with a negative O'Connell effect (Max I is lower than Max II). The other K-type contact binaries e.g., AD Cnc (Qian et al. 2007); BI Vul (Qian et al. 2013); CSTAR 038663 (Qian et al. 2014) with a deep convective envelope and fast rotation can produce a strong magnetic dynamo and solar-like activity i.e., photospheric cool spots. In this way, asymmetry in light curves can be explained as a result of spot activities. Thus, we add spot in both data sets to get a better fit. In W-D code, a spot has four parameters: the latitude ($\theta$) and longitude ($\psi$) of spot center in degrees, spot angular radius ($r_{s}$) in radians and spot temperature factor (the ratio between the spot temperature and the photosphere surface temperature of the star). In addition, we also add the third light ($l_{3}$) as an adjustable parameter in order to get a better fit. After modeling, it is found that the third light can be detected in all bands with small values as listed in Table 2. The solutions show a large hot spot on the more massive component for the best fit in dataset 2. But for dataset 1, the solutions do not converge, it maybe because of many groups of small spots on one or both stars which the W-D code may not work or find the solution. Thus, the best solution for dataset 1 has not obtained. The observed and the synthetic light curves are plotted in Figure 8 with their corresponding parameters listed in Table 2 and their geometrical configurations at phase of 0.00, 0.25, 0.50 and 0.75 shown in Figure 9.

\begin{table}
\scriptsize
\caption{Photometric solutions for 2016 and 2017 in V, R and I light curves}
\begin{center}
\begin{tabular}{lcccc}\hline
Parameters & 2016  & 2017 & 2017 & 2017\\
(models)   &(unspotted)  &(unspotted) &(l3) &(spot+l3)\\
\hline
$T_1(K)$ & 4908&assumed&assumed&assumed\\
$g_1=g_2$& 0.32&assumed&assumed&assumed\\
$A_1=A_2$& 0.50 &assumed&assumed&assumed\\
\\
$q$ &2.923($\pm0.049$)  &2.736($\pm0.030$) &2.569($\pm0.057$) &2.635($\pm0.043$)\\
$T_2(K)$ &4557($\pm9$)&4675($\pm8$) &4674($\pm8$) &4658($\pm7$)\\
$T_1 - T_2 (K)$ &351 &233 &234 &250\\
$i(^o)$ &82.528($\pm0.453$) &82.943($\pm0.358$) &83.835($\pm0.555$) &83.321($\pm0.346$)\\
$\Omega_{in}$  &2.5575  &2.6067 &2.6561 &\\
$\Omega_{out}$ &2.3462  &2.3821 &2.4181 &\\
$\Omega_1=\Omega_2$ &6.4139($\pm0.0626$) &6.2051($\pm0.0381$) &5.9682($\pm0.0829$) &6.0698($\pm0.0078$)\\
\\
$L_1/(L_1+L_2)$(V) &0.4545($\pm0.0035$) &0.3538($\pm0.0022$) & \\
$L_1/(L_1+L_2)$(R) &0.4362($\pm0.0029$) &0.3391($\pm0.0019$) & \\
$L_1/(L_1+L_2)$(I) &0.4545($\pm0.0035$) &0.3302($\pm0.0018$) & \\
\\
$L_1/(L_1+L_2+L_3)$(V) &&&0.3544($\pm0.0004$) &0.3586($\pm0.0006$)\\
$L_1/(L_1+L_2+L_3)$(R) &&&0.3461($\pm0.0004$) &0.3497($\pm0.0005$)\\
$L_1/(L_1+L_2+L_3)$(I) &&&0.3322($\pm0.0003$) &0.3349($\pm0.0004$)\\
\\
$L_3/(L_1+L_2+L_3)$(V) &&&0.0364($\pm0.0024$) &0.0233($\pm0.0013$)\\
$L_3/(L_1+L_2+L_3)$(R) &&&0.0191($\pm0.0046$) &0.0038($\pm0.0033$)\\
$L_3/(L_1+L_2+L_3)$(I) &&&0.0333($\pm0.0023$) &0.0185($\pm0.0012$)\\
\\
$r_1(pole)$ &0.2831($\pm0.0014$) &0.2805($\pm0.0008$) & 0.2856($\pm0.0023$) &0.2829($\pm0.0006$)\\
$r_1(side)$ &0.2963($\pm0.0016$) &0.2929($\pm0.0009$) & 0.2985($\pm0.0025$) &0.2956($\pm0.0007$)\\
$r_1(back)$ &0.3358($\pm0.0021$) &0.3287($\pm0.0012$) & 0.3349($\pm0.0030$) &0.3315($\pm0.0012$)\\
$r_2(pole)$ &0.4361($\pm0.0052$) &0.4421($\pm0.0034$) & 0.4408($\pm0.0076$) &0.4413($\pm0.0006$)\\
$r_2(side)$ &0.4652($\pm0.0069$) &0.4733($\pm0.0046$) & 0.4721($\pm0.0103$) &0.4725($\pm0.0007$)\\
$r_2(back)$ &0.4896($\pm0.0092$) &0.5005($\pm0.0062$) & 0.5008($\pm0.0141$) &0.5006($\pm0.0009$)\\
\\
Lat(deg)&&&&95(assumed)\\
Long(deg)&&&&255(assumed)\\
Radius(deg)&&&&89.185($\pm9.942$)\\
Temp. factor&&&&1.0041($\pm0.0004$)\\
Spot Temp. (K)&&&&4677($\pm7$)\\
\\
$f$ &16.2$\%$($\pm5.5\%$) &9.8$\%$($\pm3.1\%$) &11.6$\%$($\pm6.8\%$) &9.7$\%$($\pm1.3\%$)\\
$\Sigma{W(O-C)^2}$ &0.00311 &0.00177 &0.00149 &0.00143\\
\hline
\end{tabular}
\end{center}
\end{table}

\section{Orbital period investigation}

\begin{figure}
\begin{center}
\includegraphics[angle=0,scale=0.3]{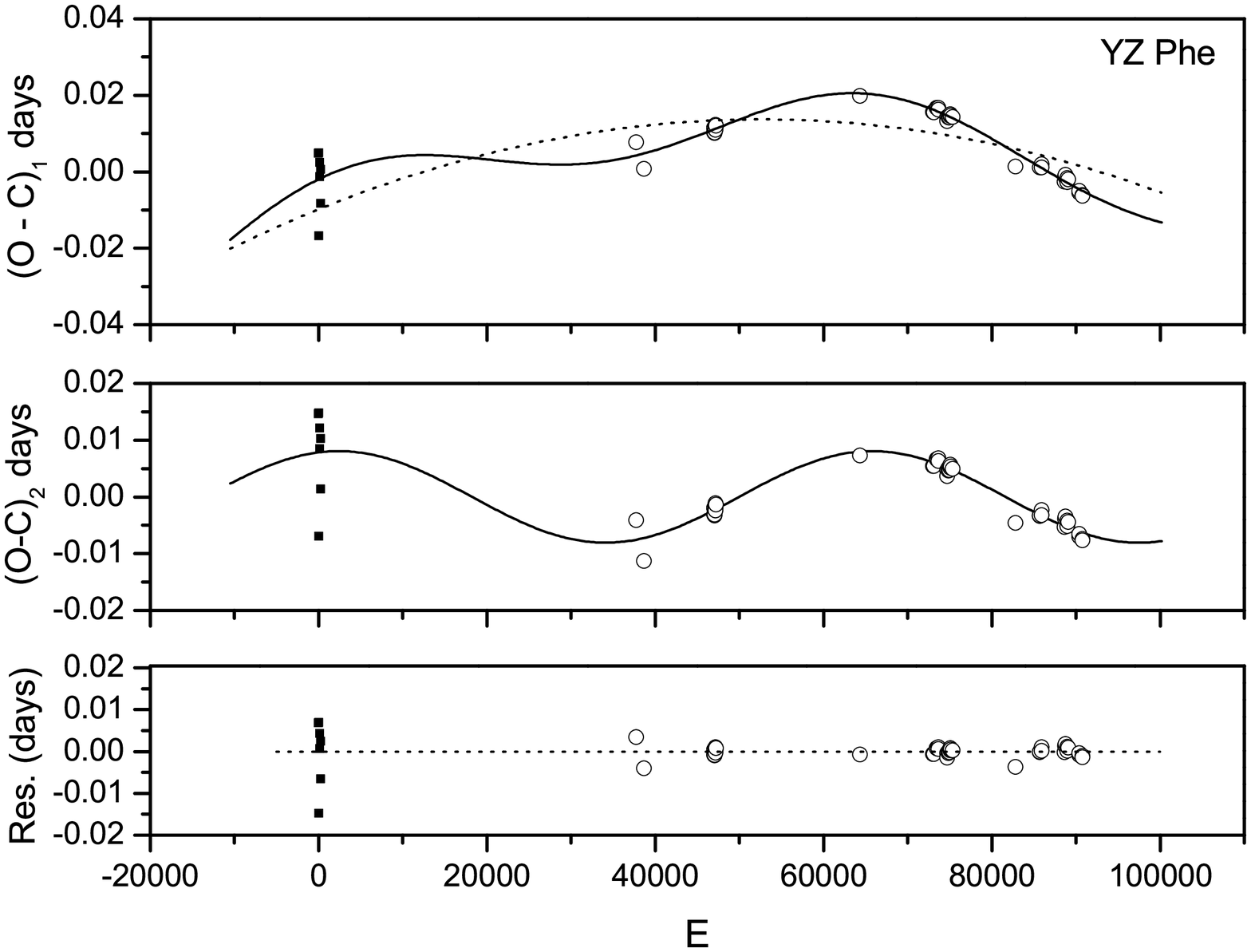}
\caption{The $(O-C)_1$ diagram in the upper panel is constructed by using the linear ephemeris in Eq. (1). The solid line in the top panel refers to a combination of long-term period decrease and cyclic change, while the dashed line refers to a downward parabolic curve which showing a secular period decrease. The seven dark- squares refer to the photographic data (pg) with weight 1 and the open circles refer to the photoelectric (pe) and ccd data with weight 8, respectively.}
\end{center}
\end{figure}

Earlier epoches and $O-C$ data of YZ Phe were published by many authors e.g., Gessner \& Meinunger (1974); Jones (1989); Kilkenny \& Marang (1990), and recently by Samec \& Terrell (1995). Although YZ Phe has been investigated for more than 50 years but no variation of orbital period change was found, rather the orbital period study reported that its orbital period remained constant at 0.23472 days for over 30 years (Samec \& Terrell, 1995). However, the period study of YZ Phe has been neglected for more than 20 years since last publication in 1995. Based on our photometric observations , new times of light minimum were determined and the orbital period change was re-analysed by using $O-C$ (observed minus calculated) method. In order to re-investigate the orbital period change of YZ Phe, all available times of light minimum from the literature were compiled together with our data. The $O - C$ data in Table 3 were recomputed with linear ephemeris given by Kreiner (2004):
\begin{equation}
Min.I (HJD) = 2436765.622+0^{d}.23472665E
\end{equation}

The result is shown in the upper panel of Figure 10, the ephemeris of YZ Phe is needed to be revised due to a downward parabolic pattern and a cyclic change. To get a satisfied fit for the trend of $(O-C)_1$ curve, it has to combine a quadratic term and sinusoidal term by assuming that the oscillation is circular ($e$ = 0). All times of minimum light are listed in Table 3 and the details of $O-C$ analysis are plotted in Figure 10. By using a least-squares method, the new ephemeris is determined as:

\begin{equation}
\begin{array}{lll}
Min.I(HJD) =  2436765.61217(\pm 0.00038)
\\ +0.23472754(\pm 0.00000001)E
\\ -[8.48(\pm 0.14)\times 10^{-12}]E^2
\\ +0.0081(\pm 0.0002) \times \sin[0^{\circ}.00568E+76^{\circ}.549(\pm 1^{\circ}.238)]
\end{array}
\end{equation}

According to Eq. (2), the sinusoidal term suggests that it has an oscillation with a period of 40.76 years and a small amplitude of cyclic variation of 0.0081 days. The $(O-C)_2$ values without the quadratic term are plotted in the middle panel and the residuals are plotted in the lowest panel of Figure 10, respectively. Although the $(O-C)_1$ data do not cover a whole cycle and there were no eclipse timings observed between Epoch = 0 to Epoch = 40000, but most of $O - C$ data are well constrained with the period oscillation as shown in the middle panel of Figure 10. This may suggest that the period change is reliable. The quadratic term in Eq. (2) indicates a secular period decrease at a rate of $\mathrm{d}P/\mathrm{d}t = -2.64(\pm 0.04)\times10^{-8}$ d/yr.

\begin{table}
\scriptsize
\caption{Times of minimum light for YZ Phe.}
\begin{center}
\begin{tabular}{lcrrccc}\hline
HJD(2400000+) & Error(days) & E & $(O-C)_1$ & Method & Min & Ref. \\
\hline
36758.5850	&		&	-30.0	&	0.00480	&	pg	&	I	&	(1)	\\
36764.5490	&		&	-4.5	&	-0.01673	&	pg	&	II	&	(1)	\\
36765.6270	&		&	0.0	&	0.00500	&	pg	&	I	&	(1)	\\
36787.4540	&		&	93.0	&	0.00242	&	pg	&	I	&	(1)	\\
36792.4970	&		&	114.5	&	-0.00120&	pg	&	II	&	(1)	\\
36821.6050	&		&	238.5	&	0.00069	&	pg	&	II	&	(1)	\\
36822.5350	&		&	242.5	&	-0.00821&	pg	&	II	&	(1)	\\
45621.3968	&0.0002	&	37728.0	&	0.00775	&	pe	&	I	&	(2)	\\
45836.7515	&		&	38645.5	&	0.00075	&	pe	&	II	&	(7)	\\
47792.6220	&		&	46978.0	&	0.01144	&	pe	&	I	&	(3)	\\
47793.4425	&		&	46981.5	&	0.01039	&	pe	&	II	&	(3)	\\
47793.5610	&		&	46982.0	&	0.01153	&	pe	&	I	&	(3)	\\
47794.6160	&		&	46986.5	&	0.01026	&	pe	&	II	&	(3)	\\
47803.5355	&		&	47024.5	&	0.01015	&	pe	&	II	&	(3)	\\
47807.5260	&		&	47041.5	&	0.01029	&	pe	&	II	&	(3)	\\
47832.6428	&0.0007 &	47148.5	&	0.01134	&	pe	&	II	&	(4)	\\
47833.8166	&0.0004 &	47153.5	&	0.01151	&	pe	&	II	&	(4)	\\
47834.7557	&0.0013 &	47157.5	&	0.01170	&	pe	&	II	&	(4)	\\
47836.5165	&		&	47165.0	&	0.01205	&	pe	&	I	&	(3)	\\
47836.7515	&0.0005 &	47166.0	&	0.01233	&	pe	&	I	&	(4)	\\
47837.3370	&		&	47168.5	&	0.01101	&	pe	&	II	&	(3)	\\
47849.4265	&		&	47220.0	&	0.01209	&	pe	&	I	&	(3)	\\
51868.77600	&		&	64343.5	&	0.01980	&	ccd	&	II	&	(7)	\\
53903.61718&	0.00033	&	73012.5	&	0.01564	&	ccd	&	II	&	(5)	\\
53921.57367&	0.00030	&	73089.0	&	0.01555	&	ccd	&	I	&	(5)	\\
54000.32560&	0.00024	&	73424.5	&	0.01668	&	ccd	&	II	&	(5)	\\
54008.54095&	0.00024	&	73459.5	&	0.01660	&	ccd	&	II	&	(5)	\\
54046.33212&	0.00020	&	73620.5	&	0.01679	&	ccd	&	II	&	(5)	\\
54054.31236&	0.00019	&	73654.5	&	0.01631	&	ccd	&	II	&	(5)	\\
54294.55308&	0.00018	&	74678.0	&	0.01431	&	ccd	&	I	&	(5)	\\
54295.60828&	0.00025	&	74682.5	&	0.01324	&	ccd	&	II	&	(5)	\\
54330.58348&	0.00016	&	74831.5	&	0.01417	&	ccd	&	II	&	(5)	\\
54332.57907&	0.00018	&	74840.0	&	0.01459	&	ccd	&	I	&	(5)	\\
54353.58703&	0.00027	&	74929.5	&	0.01451	&	ccd	&	II	&	(5)	\\
54361.33271&	0.00021	&	74962.5	&	0.01421	&	ccd	&	II	&	(5)	\\
54378.58545&	0.00018	&	75036.0	&	0.01455	&	ccd	&	I	&	(5)	\\
54390.55707&	0.00023	&	75087.0	&	0.01511	&	ccd	&	I	&	(5)	\\
54391.37826&	0.00029	&	75090.5	&	0.01475	&	ccd	&	II	&	(5)	\\
54445.36496&	0.00015	&	75320.5	&	0.01432	&	ccd	&	II	&	(5)	\\
56199.69904	&	0.00030	&	82794.5	&	0.00142	&	ccd	&	II	&	(6)	\\
56875.12470	&	0.00130	&	85672.0	&	0.00114	&	ccd	&	I	&	(7)	\\
56878.17610	&	0.00110	&	85685.0	&	0.00109	&	ccd	&	I	&	(7)	\\
56880.28870	&	0.00150	&	85694.0	&	0.00115	&	ccd	&	I	&	(7)	\\
56936.62389	&	0.00010	&	85934.0	&	0.00195	&	ccd	&	I	&	(7)	\\
56936.74038	&	0.00010	&	85934.5	&	0.00108	&	ccd	&	II	&	(7)	\\
57567.91670	&	0.00010	&	88623.5	&	-0.00257&	ccd	&	II	&	(6)	\\
57585.87465	&	0.00010	&	88700.0	&	-0.00121&	ccd	&	I	&	(6)	\\
57591.86057	&	0.00020	&	88725.5	&	-0.00081&	ccd	&	II	&	(6)	\\
57647.72365	&	0.00020	&	88963.5	&	-0.00268&	ccd	&	II	&	(6)	\\
57647.84202	&	0.00020	&	88964.0	&	-0.00167&	ccd	&	I	&	(6)	\\
57666.73719	&	0.00007	&	89044.5	&	-0.00200&	ccd	&	II	&	(6)	\\
57983.73210	&	0.00016	&	90395.0	&	-0.00543&	ccd	&	I	&	(6)	\\
57983.84996	&	0.00011	&	90395.5	&	-0.00493&	ccd	&	II	&	(6)	\\
58069.99350	&	0.00080	&	90762.5	&	-0.00607&	ccd	&	II	&	(7)	\\
58070.11060	&	0.00020	&	90763.0	&	-0.00633&	ccd	&	I	&	(7)	\\
\hline
\end{tabular}
\end{center}
{\footnotesize Notes.} \footnotesize (1) Gessner \& Meinunger 1976, (2) Jones 1989, (3) Kilkenny \& Marang 1990, (4) Arruda et al. 1993, (5) SWASP 2006-2007, (6) present paper, (7) http://var.astro.cz/ocgate.
\end{table}
 
In addition, we also consider updating ephemeris which is obtained from our observed eclipse times and compare the result to the linear ephemeris in Eq.(1) by Kreiner (2004). The present ephemeris from our  observed times of light minimum is derived as;
\begin{equation}
Min.I (HJD) = 2457585.87465+0^{d}.23472665E
\end{equation}

The plots of the results from our ephemeris are displayed in Figure 11 and the $O-C$ data from Eq.(3) are listed in Table 4. The pattern of $O-C$ curve between previous ephemeris and our ephemeris is not changed. We also obtain another ephemeris similar to Eq.(2), which using a least-squares method to determine as a result shown in Eq.(4). The parameters of cyclic change from Eq.(4) are determined and the results are the same as Eq.(2).

\begin{equation}
\begin{array}{lll}
Min.I(HJD) =  2457585.87850(\pm 0.00024)
\\ +0.23472604(\pm 0.00000001)E
\\ -[8.48(\pm 0.14)\times 10^{-12}]E^2
\\ +0.0081(\pm 0.0002) \times \sin[0^{\circ}.00568E+220^{\circ}.114(\pm 1^{\circ}.293)]
\end{array}
\end{equation}

\begin{figure}
\begin{center}
\includegraphics[angle=0,scale=0.3]{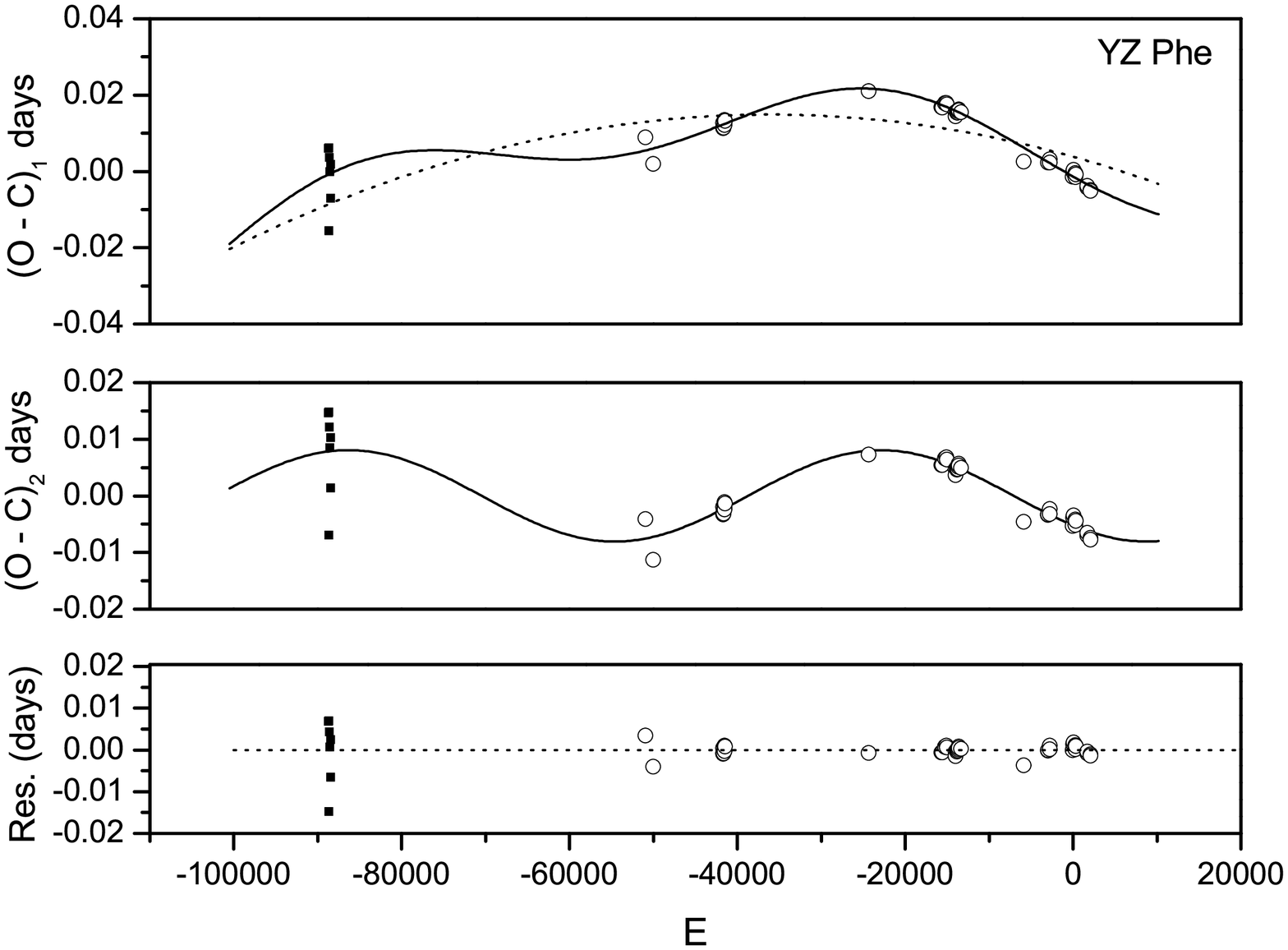}
\caption{The $(O-C)_1$ diagram in the upper panel is constructed by using new ephemeris in Eq. (3). The method and explanation are the same in Fig. 10. The analysis results do not change compared with Fig. 10.}
\end{center}
\end{figure}

\begin{table}
\scriptsize
\caption{The $O-C$ data from present ephemeris in Eq.(3)}
\begin{center}
\begin{tabular}{lcrrccc}\hline
HJD(2400000+) & Error(days) & E & $(O-C)_1$ & Method & Min & Ref. \\
\hline
36758.5850	&		&	-88730.0	&	0.00600	&	pg	&	I	&	(1)	\\
36764.5490	&		&	-88704.5	&	-0.01553	&	pg	&	II	&	(1)	\\
36765.6270	&		&	-88700.0	&	0.00621	&	pg	&	I	&	(1)	\\
36787.4540	&		&	-88607.0	&	0.00363	&	pg	&	I	&	(1)	\\
36792.4970	&		&	-88585.5	&	0.0	&	pg	&	II	&	(1)	\\
36821.6050	&		&	-88461.5	&	0.00190	&	pg	&	II	&	(1)	\\
36822.5350	&		&	-88457.5	&	-0.00701	&	pg	&	II	&	(1)	\\
45621.3968	&0.0002	&	-50972.0	&	0.00895	&	pe	&	I	&	(2)	\\
45836.7515	&		&	-50054.5	&	0.00195	&	pe	&	II	&	(7)	\\
47792.6220	&		&	-41722.0	&	0.01264	&	pe	&	I	&	(3)	\\
47793.4425	&		&	-41718.5	&	0.01160	&	pe	&	II	&	(3)	\\
47793.5610	&		&	-41718.0	&	0.01273	&	pe	&	I	&	(3)	\\
47794.6160	&		&	-41713.5	&	0.01146	&	pe	&	II	&	(3)	\\
47803.5355	&		&	-41675.5	&	0.01135	&	pe	&	II	&	(3)	\\
47807.5260	&		&	-41658.5	&	0.01150	&	pe	&	II	&	(3)	\\
47832.6428	&	0.0007	&	-41551.5	&	0.01255	&	pe	&	II	&	(4)	\\
47833.8166	&	0.0004	&	-41546.5	&	0.01271	&	pe	&	II	&	(4)	\\
47834.7557	&	0.0013	&	-41542.5	&	0.01291	&	pe	&	II	&	(4)	\\
47836.5165	&		&	-41535.0	&	0.01326	&	pe	&	I	&	(3)	\\
47836.7515	&	0.0005	&	-41534.0	&	0.01353	&	pe	&	I	&	(4)	\\
47837.3370	&		&	-41531.5	&	0.01221	&	pe	&	II	&	(3)	\\
47849.4265	&		&	-41480.0	&	0.01329	&	pe	&	I	&	(3)	\\
51868.77600	&		&	-24356.5	&	0.02100	&	ccd	&	II	&	(7)	\\
53903.61718	&	0.00033	&	-15687.5	&	0.01685	&	ccd	&	II	&	(5)	\\
53921.57367	&	0.00030	&	-15611.0	&	0.01676	&	ccd	&	I	&	(5)	\\
54000.32560	&	0.00024	&	-15275.5	&	0.01789	&	ccd	&	II	&	(5)	\\
54008.54095	&	0.00024	&	-15240.5	&	0.01781	&	ccd	&	II	&	(5)	\\
54046.33212	&	0.00020	&	-15079.5	&	0.01799	&	ccd	&	II	&	(5)	\\
54054.31236	&	0.00019	&	-15045.5	&	0.01752	&	ccd	&	II	&	(5)	\\
54294.55308	&	0.00018	&	-14022.0	&	0.01552	&	ccd	&	I	&	(5)	\\
54295.60828	&	0.00025	&	-14017.5	&	0.01444	&	ccd	&	II	&	(5)	\\
54330.58348	&	0.00016	&	-13868.5	&	0.01537	&	ccd	&	II	&	(5)	\\
54332.57907	&	0.00018	&	-13860.0	&	0.01579	&	ccd	&	I	&	(5)	\\
54353.58703	&	0.00027	&	-13770.5	&	0.01571	&	ccd	&	II	&	(5)	\\
54361.33271	&	0.00021	&	-13737.5	&	0.01541	&	ccd	&	II	&	(5)	\\
54378.58545	&	0.00018	&	-13664.0	&	0.01575	&	ccd	&	I	&	(5)	\\
54390.55707	&	0.00023	&	-13613.0	&	0.01631	&	ccd	&	I	&	(5)	\\
54391.37826	&	0.00029	&	-13609.5	&	0.01596	&	ccd	&	II	&	(5)	\\
54445.36496	&	0.00015	&	-13379.5	&	0.01552	&	ccd	&	II	&	(5)	\\
56199.69904	&	0.00030	&	-5905.5	&	0.00262	&	ccd	&	II	&	(6)	\\
56875.12470	&	0.00130	&	-3028.0	&	0.00235	&	ccd	&	I	&	(7)	\\
56878.17610	&	0.00110	&	-3015.0	&	0.00230	&	ccd	&	I	&	(7)	\\
56880.28870	&	0.00150	&	-3006.0	&	0.00236	&	ccd	&	I	&	(7)	\\
56936.62389	&	0.00010	&	-2766.0	&	0.00315	&	ccd	&	I	&	(7)	\\
56936.74038	&	0.00010	&	-2765.5	&	0.00228	&	ccd	&	II	&	(7)	\\
57567.91670	&	0.00010	&	-76.5	&	-0.00136	&	ccd	&	II	&	(6)	\\
57585.87465	&	0.00010	&	0.0	&	0.0	&	ccd	&	I	&	(6)	\\
57591.86057	&	0.00020	&	25.5	&	0.00039	&	ccd	&	II	&	(6)	\\
57647.72365	&	0.00020	&	263.5	&	-0.00147	&	ccd	&	II	&	(6)	\\
57647.84202	&	0.00020	&	264.0	&	-0.00047	&	ccd	&	I	&	(6)	\\
57666.73719	&	0.00007	&	344.5	&	-0.00079	&	ccd	&	II	&	(6)	\\
57983.73210	&	0.00016	&	1695.0	&	-0.00422	&	ccd	&	I	&	(6)	\\
57983.84996	&	0.00011	&	1695.5	&	-0.00373	&	ccd	&	II	&	(6)	\\
58069.99350	&	0.00080	&	2062.5	&	-0.00487	&	ccd	&	II	&	(7)	\\
58070.11060	&	0.00020	&	2063.0	&	-0.00513	&	ccd	&	I	&	(7)	\\
\hline
\end{tabular}
\end{center}
{\footnotesize Notes.} \footnotesize (1) Gessner \& Meinunger 1976, (2) Jones 1989, (3) Kilkenny \& Marang 1990, (4) Arruda et al. 1993, (5) SWASP 2006-2007, (6) present paper, (7) http://var.astro.cz/ocgate.
\end{table}

\section{Discussions and conclusions}

Based on light curve solutions and period analysis, we can conclude that YZ Phe is a W-subtype contact binary which the hotter component is the less massive one. The mass ratio is $q$ = 2.635($\pm0.043$) and the contact is weak with low fill-out factor of $f\sim$ 10\%. According to a spectral type of K2 V (Gaia DR2) for the hotter component, the estimated mass of $M_2$ = 0.742$M_{\odot}$ (Cox 2000). With mass ratio relation, the mass of $M_1$ = 0.281$M_{\odot}$ and $M_{total}$ $\sim$ 1.02$M_{\odot}$. Based on the mass function of the system; $f(M)=(M_1 + M_2)\sin^{3} (i)$, where inclination $i$ of the binary is $83^{\circ}.321$, thus the mass function of YZ Phe can be computed as $f(M)\sim$ 1.02$M_{\odot}$. The absolute parameters of YZ Phe can be derived; $M_{1}=0.28M_{\odot}$, $M_{2}=0.74M_{\odot}$, a=1.6112$R_{\odot}$, $R_{1}=0.4887R_{\odot}$, $R_{2}=0.7596R_{\odot}$, $L_{1}=0.1242L_{\odot}$ and $L_{2}=0.2434L_{\odot}$, respectively. Our results are  close to the parameters published by Samec \& Terrell (1995) with mass ratio around 0.4 (nearly 1/q = 0.3795 for our best fit) and a large hot spot as superluminous region (spot radius $>$ 50 deg) on the more massive component.

According to the period study in previous section, the orbital period of YZ Phe is decreasing at a rate of $\mathrm{d}P/\mathrm{d}t = -2.64(\pm 0.04)\times10^{-8}$ d /yr and the timescale by period decrease is $P/\dot{P}$ $\sim$ 8.89$\times10^{6}$ yr (or 8.89 Myr). The orbital period decrease can be explained by mass transfer or combination of mass transfer and angular momentum loss (AML) via magnetic braking. If this long-term period decrease is only due to the mass transfer from the more massive component to the less massive one, the transfer rate can be computed with the following equation:
\begin{equation}
\frac{\dot{P}}{P} = -3\dot{M_1}(\frac{1}{M_1} - \frac{1}{M_2})
\end{equation}

The mass transfer rate is $\dot{M}$ = 1.69$\times10^{-8}$ $M_{\odot}$ /yr and timescale by mass transfer can be estimated as $M_2/\dot{M}$ $\sim$ 4.39$\times10^7$ yr (or 44 Myr) and its thermal timescale by the more massive component is 9.39$\times10^7$ yr (or 94 Myr). These timescales are put together in Table 5 and it is shown that the timescale by mass transfer is taken more than timescale by observed period change via the $O-C$ curve. This may suggest that the long-term period decrease of YZ Phe may be caused by the combination of mass transfer and AML. Therefore, we can compute the decrease rate of period change by AML via magnetic braking (see Bradstreet \& Guinan 1994) in Eq.(6), where $k^2$ is the gyration constant ranging from 0.07 to 0.15 for solar type stars. By adopting a value of $k^2$ = 0.1 (Bradstreet \& Guinan 1994), the change rate due to AML can be computed as $\mathrm{d}P/\mathrm{d}t = -3.6\times10^{-8}$ d $\textrm{yr}^{-1}$ and the timescale from this way is $P/\dot{P} \sim 6.52\times10^{6} \textrm{yr}$ (or 6.52 Myr) which is close to the value by observed period change via the $O-C$ curve (8.89 Myr). This may conclude that the long-term period decrease of YZ Phe is mainly constrained by AML. It may note that the magnetic torque via stellar wind in YZ Phe is strong and causing magnetic breaking or AML. By this way, YZ Phe will become deeper contact and more evolved contact binary with small mass transfer.

\begin{table}
\caption{The computed timescales from available methods}
\begin{center}
\begin{tabular}{lcc}\hline
Methods & timescales (Myr) \\
\hline 
Period change & 8.89\\
Mass transfer & 44.0\\
Thermal & 94.0\\
\hline
\end{tabular}
\end{center}
\end{table}

\begin{equation}
\dot{P} \approx -1.1\times10^{-8}q^{-1}(1+q)^2(M_1 + M_2)^{-5/3}k^2
         \times(M_{1}R^{4}_{1} - M_{2}R^{4}_{2})P^{-7/3}
\end{equation}

According to photometric time series in Figure 4, the maxima change from time to time with cyclic behaviour and a period of 4.20 yr. This brightness modulation may involve with the longitudinal motions of spots in stellar differential rotations of host star, similar to the case of single stars e.g., Proxima Centauri (Wargelin et al. 2017) with rotational period of 83 d and magnetic cycle period of 7 yr, other rotationally variable stars e.g., Cepheid (Kiraga 2012) with long-term photometric variations. But in close binaries, there are also found the long-term variations of maximum light e.g. V1084 Sco (Pilecki et al. 2007) with period modulation of 5 yr and semi-amplitude of 15 mmag. They noted that the G-type contact binary V1084 Sco (P = 0.303 d) with spectroscopic quadruple system shows a high period decrease rate and long-term photometric modulation with a period of 5 yr, where this effect may play a crucial contribution to its long-term period decrease other than LTTE. In addition, the variations of Max differences (Max I - Max II) in Figure 6 can be interpreted as variation of the O'Connell effect with magnetic cycles of 1.28 yr. This means that sometimes Max I is lower than Max II (negative O'Connell effect), sometimes Max I is higher than Max II (positive O'Connell effect), and sometimes it has no variation of maximum (no O'Connell effect). Those variations may be caused by the variation of starspots. The O'Connell effect can be divided into two stages between active stage (high spot activity) and inactive stage (low spot activity) as pointed out by Qian et al. (2014). The intrinsic light curve variations have been found in other contact binaries e.g., EPIC 211957146 (Sriram et al. 2017); AD Cnc (Qian et al. 2007); CSTAR 038663 (Qian et al 2014); OO Aql (Li et al. 2016) and so on (see Table 6). Those systems compose of two datasets which observed in different year and all were found to have light curve variations with certain magnetic cycle lengths which differ from their periodic perturbations in the $O-C$ curves. The Applegate mechanism cannot explain those results. Thus, those contact binaries with cyclic changes in the $O-C$ curves may be interpreted as LTTE induced by the presence of a third body. The light curve asymmetries was also found in many contact binaries e.g., BI CVn (Qian et al. 2008); CW Cas (Wang et al. 2014); GN Boo (Wang et al. 2015); EQ Tao (Li et al. 2014); AB And (Djurasevic et al. 2000); BX Peg (Lee et al. 2004, 2009); BB Peg (Kalomeni et al. 2007); AA UMa (Lee et al. 2011); V410 Aur (Luo et al. 2017) and recent study of four contact binaries by Djurasevic et al. (2016). In Table 6, it can see that most of active contact binaries show small one or two cool spots on the more massive component rather than hot spot. But YZ Phe shows a large hot spot ($>$ 50 deg) in Figure 9 which differs from those systems and its spot is likely to grow from 54 deg (Sameg \& Terrell 1995) to 89 deg. The big spotted area can be formed from many groups of small spots (Suarez Mascareno et al. 2016) and each spot can vary with time, but some spots may stay for long timescale without changing e.g., CSTAR 038663 (Qian et al 2014) and PZ UMa (Zhou \& Soonthornthum 2019). This may suggest that spots and magnetic activities in contact systems are complex and the mechanism or physical processes behind the large hot spot in YZ Phe is unknown. This effect in YZ Phe may play a contribution to its long-term brightness modulation. In order to understand those effects i.e., the rotations and magnetic activity cycles in active contact binaries, it is needed the optical data from photometric time series (e.g., ASAS and SSS data) to explore the long-term photometric modulation and from high precision time of photometry (e.g., Kepler and SWASP data) to determine the shorter period of magnetic activity cycle. Spectroscopic observations and other wavelength surveys will be helpful to confirm for both rotation and stellar activity cycles in active contact binaries or even in active close binaries.

\begin{table}
\scriptsize
\caption{Some contact binaries with variations of the O'Connell effect (magnetic cycles) and LTTE.}
\begin{center}
\begin{tabular}{lcccccccc}\hline
Contact binaries  &T1 (Sp.) &Period &$q$ &dP/dt & spots& magnetic cycles & LTTE &Ref.\\
 (A- or W-subtype)& (K)& (d)& & ($\times 10^{-8}$ d/yr)& (deg)& Period (yr)& Period (yr)& \\
\hline
OO Aql (A)& 6100 (F8/G1)& 0.5068& 0.844& -3.63& unkown& 21.5& 69.3& (1)\\
V396 Mon (W)& 6210 (F8)& 0.3963& 0.392& -8.57& 1 cool (11.4)& unknown& 42.4& (2)\\
EPIC 211957146 (A)& 5950 (G)& 0.3550& 0.170& +106& 1 cool (13.5) & unknown& 16.23& (3)\\
GSC 03526 (W)& 4830 (K)& 0.2922& 0.351& decrease& 1 cool (18-22)& unknown& 7.39& (4)\\
AD Cnc (W)& 5000 (K0)& 0.2827& 0.770& +49.4& 2 cool (12-18)& 17& 6.6& (5)\\
VW Cep (W)& 5050 (K0)& 0.2783& 0.302& -16.9& 2 cool (16-23)& 7.62& & (6)\\
CSTAR 038663 (W)& 4616 (K4)& 0.2671& 1.120& inclease& 1 cool (15.5)& $>$116 d& 1 d/2 yr& (7)\\
PZ UMa (A)& 5430 (G7)& 0.2627& 0.178& inclease& 1 cool (40.0)& unknown& 13.22& (8)\\
YZ Phe (W) & 4908 (K2)& 0.2347& 0.379& -2.64& 1 hot ($>$ 50)& 1.28& 40.76& (9)\\
\hline
\end{tabular}
\end{center}
{\footnotesize Notes.} \footnotesize (1) Li et al. 2016, (2) Liu et al. 2011, (3) Sriram et al. 2017, (4) Liao et al. 2012, (5)  Qian et al. 2007, (6) Mitnyan et al. 2018, (7) Qian et al. 2014, (8) Zhou \& Soonthornthum 2019, (9) this paper
\end{table}

W UMa contact binaries consist of components with spectral type between A and K, and their orbital periods ranging from 0.2 to 1.5 days (Rucinski 1998). The late-type stars, i.e. Spec. F to K with rapid rotation are expected to have high levels of chromospheric activity and X-ray emission, e.g., VW Cep (Mitnyan et al. 2018) and 2MASS J11201034-2201340 (Hu et al. 2016). However, it has no report about X-rays fluxes and radio emission measurement related to YZ Phe (e.g., Beasley et al. 1993; Stepien et al. 2001; Chen et al. 2006). Based on the results in previous section, the cyclic oscillation ($A_3$ = 0.0081 days and $P_3$ = 40.76 yr) in $O-C$ curve can be explained either by Applegate mechanism or light-time effect via the presence of a third body. The Applegate mechanism (Applegate 1992) suggested that the cyclic change is caused by orbital period modulation when a star goes through its magnetic activity cycle in the quadrupole moment of solar-like components. Similar idea had been studied by Lanza \& Rodono (2002), who found the correlation between magnetic modulation and orbital period variation. But Tran et al. (2013) found that the cyclic variation with small amplitudes (200-300 second) in short timescale and period less than 1200 days in the $O-C$ curves for minima or maxima could be induced by starspots. Based on spot study by Tran et al (2013) and Kalimeris et al. (2002), it suggests that if the amplitude of cyclic change is high (more than 300 second or 0.0035 days) in long timescale ($>$ 4 yr) with long period oscillation (more than 1200 days or 3.29 yr), the perturbations may be induced by a third body or magnetic activity cycle. This may help to distinguish between perturbations caused by spots and by a third body or magnetic cycle in the $O-C$ curves. The second idea about third body is supported by many observational evidences e.g., Liao and Qian (2010); Sriram et al. (2017) and recently by Li et al. (2018), who reported that there is a high frequency of having a third companion in contact binary when orbital period of contact binary is shorter than 0.3\,d with 65\% in statistical study. This suggests that the contact binaries with periods close to the period limit 0.22\,d, are commonly accompanied by relatively close third body. Furthermore, the period of cyclic change is long with 40.76 yr, while the period of variation of maxima from rotational motion of spots in Figure 4 is only 4.20 yr and the period of variation of the O'Connell effect in Figure 6 is more shorter with 1.28 yr. In this case, the Applegate mechanism via magnetic cycle cannot explain the cyclic variation in Figure 10 or 11. In addition, the amplitude and period of cyclic change are larger than the values or effect induced by starspots, i.e. 0.0035 days for amplitude and 3.29 yr for period. Moreover, the studies of Kepler binaries by Borkovits et al. (2015, 2016) have shown that the quasi-sinusoidal perturbations in the $O-C$ curves can be interpreted as dynamical effect or LTTE resulted by the presence of a close-in third body (Irwin 1952). The tertiary companions are believed to drive the formation and evolution of the inner close binaries as discussed by Eggleton \& Kiseleva (2006). Thus, the plausible reason to cause the cyclic perturbation in the $O-C$ curve of YZ Phe is the light-travel time effect via the presence of a third body, which is in agreement with the third light we detected in the light curve modeling  as shown in Table 2. In section photometric solutions, the modeling curves fit well when including third light and spot together. This may be the first time to report about detection of a third body for YZ Phe which confirmed by period and light curve analysis. 

Therefore, if the cyclic perturbation in Figure 10 is a plausible result of the light-travel time effect via the presence of a third body orbiting around the inner binary, the parameters of the tertiary can be computed by the mass function of it when assuming that the third body moves in a circular orbit. 
\begin{equation}
f(m) = (m_3 \sin i')^3 /(m_1 + m_2 + m_3)^2,
\end{equation}
where $f(m)$ = $\frac{4\pi}{GP^2} \times (a'_{12} \sin i')^3$. We also used $a'_{12}$ sin $i' = A_3 \times c$, where $A_3$ is the amplitude of the $O-C$ oscillation, $c$ is the speed of light and $i'$ is the inclination of the third body's orbit. Thus, the mass function $f(m) = 0.0017M_{\odot}$ for the tertiary, the mass and the orbital radius of the tertiary can be determined. The corresponding values are displayed in Table 7.

\begin{table}
\scriptsize
\caption{Parameters of the third body}
\begin{center}
\begin{tabular}{lrcl}\hline\hline
 Parameters & Value & Error & Units\\
\hline
$A_3$ &0.0081 &0.0002 & days\\
$P_3$ &40.759  &0.001 & years\\
$a'_{12} \sin{i'}$ & 1.403 & 0.035 & au\\
$f(m_3)$ & 0.0017 & 0.0001 & $M_{\odot}$\\
$e_3$ & 0.0 & assumed & -\\
$m_3$ ($i'=90^{\circ}$) & 0.130 & 0.003 & $M_{\odot}$\\
$a_3$ ($i'=90^{\circ}$) & 11.0 & 0.4 & au\\
\hline
\end{tabular}
\end{center}
\end{table}

The minimum mass of the tertiary ($i' = 90^{o}$) is estimated as $m_3$ = 0.130$M_{\odot}$ with distance $\sim$ 11 au from the central system, it may be a cool stellar companion. If the tertiary is really existent, it may play an important role for the formation and evolution of the binary system by removing angular momentum from the central binary via Kozai cycle (Kozai 1962) or a combination of Kozai cycle and tidal friction, that causes the system to have an initial short orbital period (i.e. initially detached system evolves into the present contact configuration). In addition, the long-term period decrease with mass ratio $<$ 0.4 indicates that it is in agreement with Qian (2001, 2003) that this system is on AML-controlled stage and it will evolve into a deeper contact binary via magnetic stellar wind.

\begin{acknowledgements}
We would like to thank the Chinese Natural Science Foundation (No. 11703080) for the partial support on the research grant under the research project. This paper has made use of data from the DR1 of the WASP data (Butters et al. 2010) as provided by the WASP consortium, and the computing and storage facilities at the CERIT Scientific Cloud, reg. no CZ. 1.05/3.2.00/08.0144 which is operated by Masaryk University, Czech Republic. This work has made use of data from the European Space Agency (ESA) mission {\it Gaia} (\url{https://www.cosmos.esa.int/gaia}), processed by the {\it Gaia} Data Processing and Analysis Consortium (DPAC, \url{https://www.cosmos.esa.int/web/gaia/dpac/consortium}). Funding for the DPAC has been provided by national institutions, in particular the institutions participating in the {\it Gaia} Multilateral Agreement. This research has also made use of the SIMBAD online database, operated at CDS, Strasbourg, France, NASA's Astrophysics Data System (ADS). Finally, we would like to thank Dr. Wiphu Rujopakarn and NARIT for time allocation to use PROMPT-8 for our observations, as well as Dr. Puji Irawati for other photometric data from PROMPT-5. We thank Dr. N.-P., Liu for helpful comment and discussion during this work.
\end{acknowledgements}

\end{document}